# Scalable construction of hybrid quantum photonic cavities

Andrew S. Greenspon,[1,2,7*] Mark Dong,[1,2,8*] Ian Christen,[2] Gerald Gilbert,[4] Matt Eichenfield,[3,5] and Dirk Englund[2,6,9]

[1]*The MITRE Corporation, 202 Burlington Road, Bedford, Massachusetts 01730, USA*
[2]*Research Laboratory of Electronics, Massachusetts Institute of Technology, Cambridge, Massachusetts 02139, USA*
[3]*Sandia National Laboratories, P.O. Box 5800 Albuquerque, New Mexico, 87185, USA*
[4]*The MITRE Corporation, 200 Forrestal Road, Princeton, New Jersey 08540, USA*
[5]*College of Optical Sciences, University of Arizona, Tucson, Arizona 85719, USA*
[6]*Brookhaven National Laboratory, 98 Rochester St, Upton, New York 11973, USA*
[7]*agreenspon@mitre.org*
[8]*mdong@mitre.org*
[9]*englund@mit.edu*
*\*equal contributors*

**Abstract**

Nanophotonic resonators are central to numerous applications, from efficient spin-photon interfaces to laser oscillators and precision sensing. A leading approach consists of photonic crystal (PhC) cavities, which have been realized in a wide range of dielectric materials. However, translating proof-of-concept devices into a functional system entails a number of additional challenges, inspiring new approaches that combine: resonators with wavelength-scale confinement and high quality factors; scalable integration with integrated circuits and photonic circuits; electrical or mechanical cavity tuning; and, in many cases, a need for heterogeneous integration with functional materials such as III-V semiconductors or diamond color centers for spin-photon interfaces. Here we introduce a concept that generates a finely tunable PhC cavity at a select wavelength between two heterogeneous optical materials whose properties satisfy the above requirements. The cavity is formed by stamping a hard-to-process material with simple waveguide geometries on top of an easy-to-process material consisting of dielectric grating mirrors and active tuning capability. We simulate our concept for the particularly challenging design problem of multiplexed quantum repeaters based on arrays of cavity-coupled diamond color centers, achieving theoretically calculated unloaded quality factors of $10^6$, mode volumes as small as $1.2(\lambda/n_{eff})^3$, and maintaining >60% total on-chip collection efficiency of fluorescent photons. We further introduce a method of low-power piezoelectric tuning of these hybrid diamond cavities, simulating optical resonance shifts up to ~760 GHz and color center fluorescence tuning of 5 GHz independent of cavity tuning. These results will motivate integrated photonic cavities toward larger scale systems-compatible designs.

**Introduction**

Photonic crystal (PhC) cavities are ubiquitous in optics, with applications including modulation [1], spectroscopy [2], filtering [3,4], sensing [5,6], lasers, nonlinear optics [7], and coupling to quantum emitters [8]. Particularly in quantum information processing, PhC cavities are useful for improving optical coupling to atomic systems that include color centers in diamond [9–12] and silicon carbide [13], quantum dots in III-V semiconductors [14,15], or other atoms in other solid-state systems [16,17]. Moreover, widely tunable cavity resonances with stable embedded quantum emitters simplifies the practical generation of identical photons for remote-entanglement operations necessary for creating quantum networks or network-based quantum computers [18]. Monolithic diamond PhC cavities have been fabricated with high quality factor [19], demonstrating Purcell enhancement [20] of the emission from implanted nitrogen vacancies (NVs) [21] and group-IV color centers [9,10,22]. Other reported plasmonic-photonic cavities [23] for near-surface NVs have a simulated mode volume of ~0.1 $(\lambda/n)^3$. Additional hybrid cavities have been constructed by patterning high index cavities, usually microdisk, that are then transferred to bulk diamond [24–26].

However, scalable and repeatable fabrication of nanoscale features to define cavities remains difficult, especially in materials that either lack mature processing techniques, are limited to small substrates, or when short optical wavelengths are required. Some difficulties in diamond cavities include color centers placed too close to nearby patterned surfaces, which can have lattice damage from plasma etching [27] or create unwanted surface state dynamics [28,29]. Variations in surface states and other defect states may cause ionization of the color centers from the spin-active negatively charged state to a neutral state [30], along with spectral broadening and diffusion [31], and decrease in spin coherence time [32,33]. Moreover, cavities composed of non-CMOS-compatible dielectrics can be difficult to integrate into large-scale, CMOS-fabricated photonic integrated circuits (PICs) and may require optical isolation from the substrate to avoid low quality factor and high losses into the substrate. Optimal placement of the emitter at the high-field location to optimize Purcell enhancement is a stochastic process [11].

In this work, we propose a hybrid PhC cavity concept that addresses several challenges of scalable cavity fabrication. The design consists of two parts: 1) a functional dielectric or semiconductor (III-V or quantum material) patterned into a simple waveguide and 2) a CMOS-fabricated target substrate with high-resolution grating mirrors and cavity tuning capabilities made with standardized processing techniques. By heterogeneously stamping material 1 on material 2, we not only improve the overall fabrication precision but enable a tailorable cavity with an optical mode based on pre-selected devices. We consider in particular the application to cavity-enhanced spin-photon interfaces in quantum information processing applications. We show through simulations that by simply placing color centers in diamond waveguide "quantum microchiplets" (QMCs) onto a patterned silicon nitride (SiN) nanostructure, we can form hybrid PhC cavities with high quality factors ($Q > 10^6$) and small mode volumes ($V \sim 1.2\ (\lambda/n_{eff})^3$). We also calculate cavity out-coupling via the diamond waveguide with theoretical Purcell enhancement of >895 with 90.5% of the light emitted into the desired optical mode for photon routing via an underlying PIC. Last, we simulate the integration of these cavities with a piezoelectrically actuated photonics platform [34], enabling low-power piezoelectric tuning of both the cavity mode (>760 GHz) and color center emission spectrum (5 GHz independent of cavity tuning, up to 50 GHz with cavity-dependent tuning) through strain and moving boundary effects. The tuning capability should improve the yield of identical spin-active emitters for high-fidelity quantum information processing.

**Hybrid cavity concepts and designs**

The general design of the PhC hybrid cavity is shown in Fig. 1. A pre-etched nanostructure containing quantum emitters, such as any dielectric or semiconductor material waveguide etc. is heterogeneously integrated via direct placement on a patterned layer stack on a silicon wafer. For our particular geometry, we assume a diamond waveguide placed (along the *x* direction) on a substrate consisting of SiN on top of a thick layer of $SiO_2$.

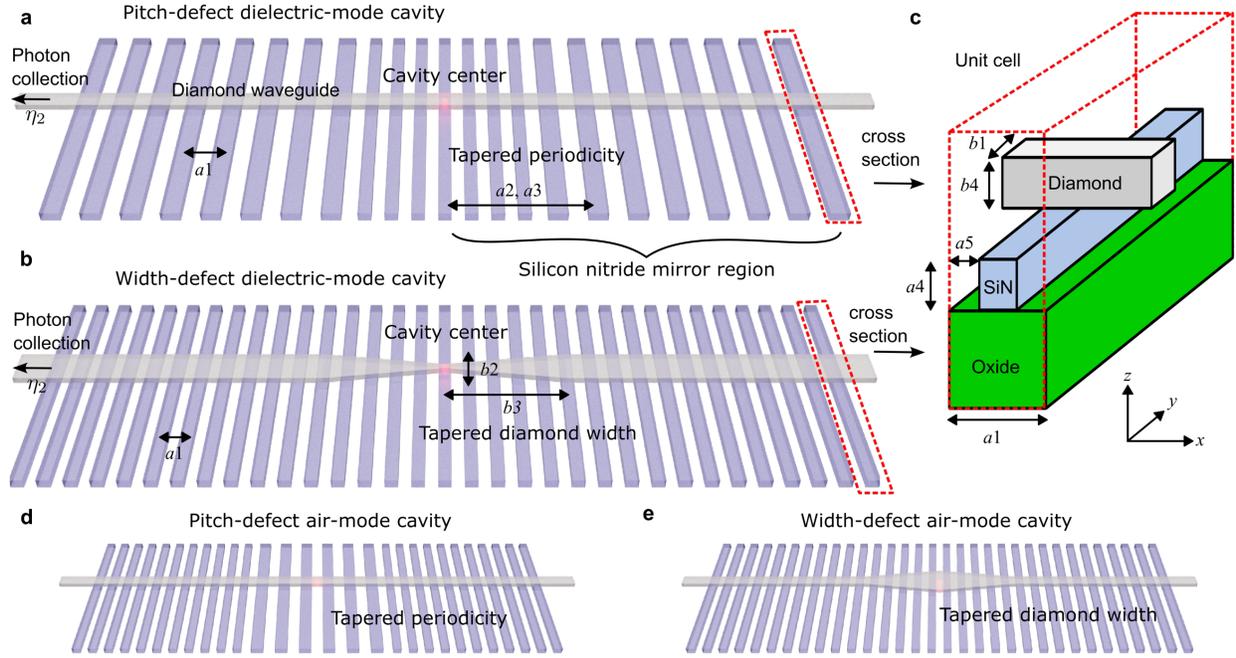

**Figure 1: General design of a hybrid cavity.** A hybrid cavity is composed of a diamond waveguide placed on top of a dielectric (silicon nitride) grating. These two structures combined form a hybrid photonic crystal cavity. a) Schematic of a pitch-defect cavity geometry for a dielectric optical mode where the SiN mirrors have a tapered periodicity starting from narrow at the cavity center to wider at the edges. b) Schematic of a width-defect cavity geometry for a dielectric optical mode where the diamond waveguide width is tapered starting from narrow at the cavity center to wider at the edges. c) Unit cell of a hybrid cavity with diamond nanobeam, patterned SiN structures and an underlying $SiO_2$ on Si substrate. The cavity modes can be tuned by multiple geometric parameters including diamond height $b4$, width $b1$, width tapering $b2$, $b3$, grating period $a1$, grating taper length $a2$, $a3$, SiN height $a4$, and grating fill factor $a5$. d)-e) Schematics of the corresponding air optical modes of pitch-defect and width-defect hybrid cavities, respectively. In these cases, the pitch-defect and waveguide widths are tapered starting from wider at the cavity center to narrower at the edges.

Confinement of the optical mode is naturally achieved by the diamond waveguide in the lateral ($y$) and vertical ($z$) directions while the evanescent field interacts with the underlying SiN. The SiN is patterned to form lines alternating with air with a specified pitch and duty cycle to create a band gap between allowed frequency bands in the hybridized diamond-SiN system — here the bands are the canonical air and dielectric modes of a periodic structure. Within the gap, no optical band is supported so the system acts as a mirror. In this hybrid architecture, a cavity is formed by varying device parameters such that a band — either air or dielectric — becomes locally allowed. Outside of this varied region, the system returns to the mirror state, leaving an island of confinement amid the mirror. In our designs, we use a parabolic [35] variation of either the diamond waveguide width (width defect) or grating pitch (pitch defect) to locally tune the band and create the cavity. There are thus four distinct cavity configurations depending on whether the width or pitch are used to create the defect and toward which allowed band the defect is tuned, each of which may be preferred depending on the physical situation: i) pitch defect dielectric mode where the pitch increases from the center, confining the optical mode in the diamond and underlying SiN regions; ii) pitch defect air mode where the pitch decreases from the center, confining the optical mode in the diamond and underlying air regions; iii) width defect dielectric mode where the diamond width increases from the center, confining the optical mode in the diamond and underlying SiN regions; iv) width defect air mode where the diamond width decreases from the center, confining the optical mode in the diamond and underlying air regions. We note that, in all designs, the underlying patterned lines have an important translation invariant symmetry in the longitudinal ($y$) direction, allowing greater tolerance in aligning the diamond waveguide during heterogeneous integration.

There are several geometric parameters that can be varied in this system for achieving the desired cavity mode. We define each of the geometric parameters in a set $\{a1, a2, a3, a4, a5, b1, b2, b3, b4\}$ with definitions given in Table 1. For a pitch defect cavity, $a2$ and $a3$ are varied while $b2$ and $b3$ are set to 0 (vice versa for a width defect). The pitch defect varies according to $a^{(j)} = a1 + A(j/a3)^2$ where $a^{(j)}$ is the pitch at the $j$th position away from a central pitch $a1$, $a3$ is the number of periods over which the pitch changes by $A$. After $a3$ periods, the remaining lines have pitch $a1 + A$ for the mirror section. For a width defect, the equation is similar save for the variation in waveguide width is given by a continuous coordinate (instead of discrete), which defines the parabolic curve variation between the central and mirror width sections.

**Table 1:** List of geometric parameters to generate hybrid cavities.

| Variable | Definition |
| --- | --- |
| $a1$ | central grating pitch |
| $a2$ | adiabatic taper length |
| $a3$ | number of periods in taper |
| $a4$ | SiN thickness |
| $a5$ | etch duty cycle |
| $b1$ | waveguide center width |
| $b2$ | waveguide taper to edge difference |
| $b3$ | length of waveguide taper |
| $b4$ | waveguide height |

For general quantum networking applications, two important cavity parameters must be optimized: the spectral efficiency $\eta_1$ and the collection efficiency $\eta_2$, which combine to make up the total efficiency $\eta_{tot} = \eta_1 \eta_2$. These efficiencies are critical for scalable processing of the emitted single photons or executing remote-entanglement protocols. Specifically, we consider the case of spin-selective photon emission from color centers in diamond. The process is generally inefficient ($\eta_1 << 1$) due to significant emission into the phonon sideband (PSB) (e.g. 0.96 for NVs [36], 0.4 for SnV [37]) instead of the zero-phonon line (ZPL). Moreover, once the fluorescent photon is emitted into the cavity, efficient out collection ($\eta_2$) through a PIC or free-space for processing becomes critical. The cavity must be maximized for both parameters. The spectral efficiency $\eta_1$ is defined as [38]:

$$\eta_1 = F/(F + \frac{\Gamma_{total,0}}{\Gamma_{ZPL,0}} - 1) \qquad (1)$$

where $F = \frac{3}{4\pi}(\lambda/n_{eff})^3 (Q/V)$ is the Purcell factor of the ZPL emission at a wavelength $\lambda$ within a cavity of quality factor $Q$, mode volume $V$, and effective refractive index $n_{eff}$, $\Gamma_{total,0}$ is the emitter's total emission rate without cavity enhancement, and $\Gamma_{ZPL,0}$ is the emission rate into the ZPL absent cavity enhancement. The straightforward strategy is simply to maximize the cavity $Q$ while reducing the mode volume $V$ to improve spectral efficiency. However, adjusting the the cavity geometry to increase collection efficiency $\eta_2$ generally decreases quality factor $Q$, as the emitter photons cycle less in the cavity before exiting the desired collection path. For a cavity with a very large number of mirror sections on both sides, $Q$ will be high (with equivalent high F and $\eta_1$ approaching 1), but outcoupling will suffer. The key design challenge is to proportionally decrease the mirror reflectivity (and hence the cavity $Q$) to increase outcoupling $\eta_2$ while maintaining a high spectral efficiency $\eta_1$ to maximize the figure of merit (FoM) $\eta_{tot}$ [39]. To estimate an acceptable value of the Purcell factor that will enable a high $\eta_1$, we insert the Debye-Waller factors (0.04, 0.7, and 0.6 for the NV-, SiV-, and SnV- respectively) into Equation 1. These inputs lead to total $\frac{\Gamma_{total,0}}{\Gamma_{ZPL,0}} - 1$ values of 24, 0.43, and 0.67 respectively. Therefore, a relatively small Purcell Factor of even 10 can bring $\eta_1$ to 96% and 94% for the SiV- and SnV- (for NV- a Purcell >100 is required). By assuming a typical normalized mode volume for PhC cavities on the order of $V \sim 1.0 \ (\lambda/n_{eff})^3$, we find that cavity $Q$s need only reach ~100-1000 to maintain a high spectral efficiency. Finally, by adjusting the cavity mirrors appropriately, we then optimize the outcoupling efficiency $\eta_2$.

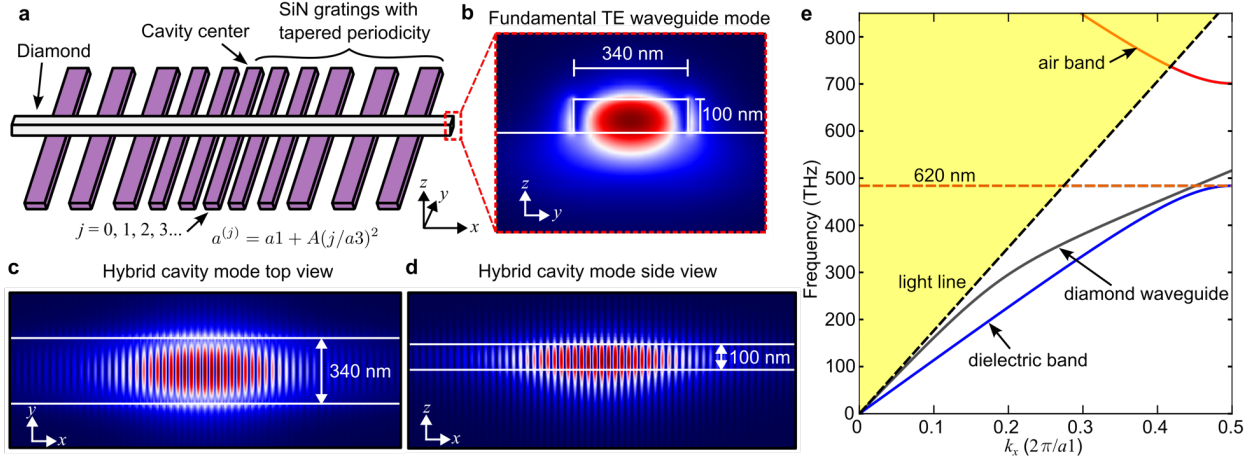

**Figure 2: Design of the pitch-defect hybrid cavity.** a) The pitch-defect hybrid cavity is composed of a diamond waveguide placed on top of a dielectric (SiN) grating with a parabolic-tapered periodicity from the center of the cavity. b) Numerical calculation of the fundamental TE diamond-SiN hybrid optical mode. c)-d) Finite-difference calculations of the hybrid cavity mode seen from the top and side views, respectively. e) Individual band diagrams for the SiN periodic grating structure (labeled dielectric and air bands) and the diamond waveguide dispersion (labeled diamond waveguide) for an exemplary hybrid cavity with $a1$=170 nm, $a5$ = 50% duty cycle, $b1$=340 nm, $b4$=100 nm.

**Hybrid cavity optical simulation results**

We proceed to design and simulate optimal cavity designs for diamond color centers using several 3D numerical simulation tools. We generally follow the design philosophy of erring on the side of nanofabrication simplicity. Thus, because repeatable fabrication of a parabolic shape on diamond waveguides may still be difficult relative to high-fidelity patterning of SiN, we focus on pitch-defect cavity designs for the remainder of this paper. Fig. 2a illustrates important design parameters for the pitch-defect cavity, while Fig. 2b-d shows a numerically calculated mode (ANSYS Lumerical) for an exemplary hybrid cavity geometry at 619.5 nm to match the SnV ZPL wavelength. We also calculated the dispersion curves (MIT Photonic Bands) for the diamond waveguide and the SiN/air grating individually, plotted in Fig. 2e. Throughout all subsequent numerical simulations, we use the following refractive indices for air $n_{air}$ = 1, SiN $n_{SiN}$ = 1.95, and diamond $n_{dmd}$ = 2.42.

The dispersion curves in Fig. 2e give initial guidance and insight for how the hybrid cavity is formed. First, the cavity frequency of interest must lie in the bandgap of the SiN/air mirror regions but outside the bandgap in the pitch-defect region. This is easily done by adjusting the pitch $a1$ appropriately. Second, the diamond waveguide mode must couple efficiently to the underlying grating structure, suggesting spatial overlap and phase matching conditions. Specifically, the approximate phase matching condition should be satisfied:

$$n_{eff,dmd} \approx (1 - a_5)n_{air} + a_5 n_{SiN} \qquad (2)$$

As the index of diamond is much larger than the index of SiN and air, making the diamond waveguide thin enables both better phase matching and spatial overlap due to a larger vertical evanescent field. We settle on a diamond thickness of $b4$ = 100 nm and widths $b1$ around 300 nm - 400 nm, which satisfy the aforementioned requirements while being within reasonable fabrication limits[40,41].

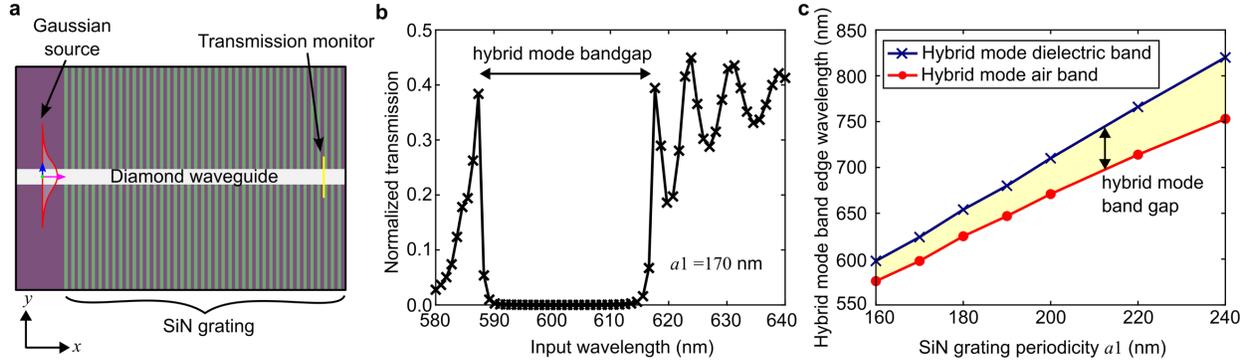

**Figure 3: Simulations of the hybrid optical mode bandgap.** a) Diagram of gaussian propagation simulations of the full 3D hybrid geometry for variable SiN periodicity and diamond waveguide dimensions of $b1$ = 340 nm wide, $b4$ = 100 nm high. b) Normalized transmission plot of the hybrid diamond-SiN propagation with $a1$ = 170 nm showing the bandgap as a function of the input wavelength, which accounts for the bandgap of the periodic structure, the diamond waveguide dispersion, and the evanescent coupling between the two materials. c) Wavelength band edges extracted from additional transmission simulations. The air and dielectric bands are plotted for various SiN periodicities.

While the band diagrams (Fig. 2e) are a good starting point, the real optical mode is hybridized between the diamond and underlying dielectric grating. We calculate the hybrid mode bandgap (Fig. 3a) using a fully 3D finite-difference wave propagation simulation (ANSYS Lumerical), which nominally follows the bands calculated in Fig. 2e. These hybrid mode simulations fully account for the phase matching and mode overlap conditions required to form a usable cavity.

With an understanding of the hybrid mode bandgap, we proceed to design a high-$Q$ quantum optical cavity. Fig. 4a shows the simulation the cavity mode is found by sweeping a TE dipole source frequency while monitoring each frequency's field decay in time. The cavity mode frequency and quality factor are extracted and plotted with respect to the diamond width (Fig. 4b, c) for different grating periodicities ($a1$). The simulated unloaded quality factor for these cavities generally ranges from $10^5$ to $10^6$ with relatively little sensitivity to different diamond geometries. Thus we see that, for an already fabricated grating mirror geometry, we can readily tune the cavity mode over tens of nanometers by selectively stamping a diamond waveguide of appropriate width. This can be useful for cavity mode adjustment as diamond waveguides may be stamped, tested, removed, and subsequently replaced, enabling coarse tuning of the cavity without resorting to new fabrication.

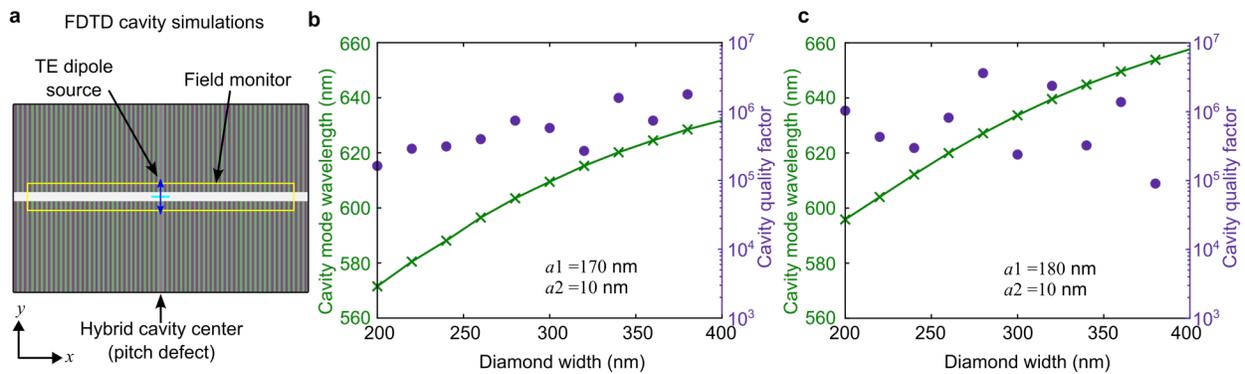

**Figure 4: Cavity mode dependence on diamond waveguide width with a fixed SiN mirror geometry.** a) Diagram of the hybrid cavity simulation with a TE point dipole source and surrounding E-field monitor. b)-c) Cavity mode wavelength and quality factor vs diamond waveguide width for a pitch-defect design with $a1$ = 170 nm and $a1$ = 180 nm, respectively. The other cavity parameters are $a2$ = 10 nm, $a3$ = 20 periods, $a4$ = 150 nm, $a5$ = 50% duty cycle, and $b4$ = 100 nm.

A summary of selected simulation geometries with quality factor and mode volume is listed in Table 2, with cavity mode wavelengths located around the NV- and SnV- ZPLs. We note that simulated quality factors are generally only accurate to within an order of magnitude due to fitting the gradual exponential decay calculated from the field monitors. The extracted mode volumes of the hybrid cavity are only slightly larger than monolithic diamond nanobeam PhC cavities [9], [11],[42] while offering repeatable and flexible cavity construction.

**Table 2:** List of geometries with $Q$ optimized cavity modes

| Parameter | Geometry 1 | Geometry 2 | Geometry 3 |
|---|---|---|---|
| $a1$ | 180 nm | 180 nm | 170 nm |
| $a2$ | 10 nm | 10 nm | 10 nm |
| $a3$ | 20 | 20 | 20 |
| $a4$ | 150 nm | 150 nm | 150 nm |
| $a5$ | 50% | 50% | 50% |
| $b1$ | 310 nm | 400 nm | 340 nm |
| $b2$ | 0 | 0 | 0 |
| $b3$ | 0 | 0 | 0 |
| $b4$ | 100 nm | 100 nm | 100 nm |
| **Cavity Mode** | **636.5 nm** | **657.7 nm** | **619.5 nm** |
| $Q$ | $5.77 \times 10^7$ | $1.50 \times 10^6$ | $3.95 \times 10^5$ |
| $V$ | $4.54 \times 10^{-20}$ m³ | $5.31 \times 10^{-20}$ m³ | $4.50 \times 10^{-20}$ m³ |
| $n_{eff}$ | 1.81 | 1.85 | 1.852 |
| $V_{norm}$ $(\lambda/n_{eff})^3$ | 1.04 | 1.19 | 1.2 |

In order to complete the hybrid cavity design, we note simply maximizing the intrinsic $Q$ is not sufficient for quantum operations. As previously mentioned, a high collection efficiency $\eta_2$ is needed. We investigate the trade-off between outcoupling and intrinsic $Q$ by calculating additional cavity geometries with variable mirror periods on one side of the cavity. With a simulation setup in Fig. 4a, we vary the number of mirror periods to find how the new "loaded" quality factor $Q$ varies with $\eta_2$. For these simulations, we also introduce some fab non-idealities for comparison to give a better sense of how real devices behave. Specifically, we assume 2 orders of magnitude lower $Q$ than calculated (this presumes additional losses purely due to scattering and absorption effects) and only 50% alignment of the color center dipole and cavity-mode electric field. We label these values with the (fab) designation in the plots.

Fig. 5 shows simulations of the 619.5 nm cavity mode $Q$, Purcell factor $F$, and total collection efficiency $\eta_{tot}$ as the number of mirror periods is varied. We find that in an idealized high-$Q$ regime, the Purcell factor likewise remains high such that $\eta_1$ is always efficient and close to 1. Here the total collection efficiency $\eta_{tot}$ is simply dominated by the output collection $\eta_2$ preferring low mirror period designs. However, in real cavities that typically operate with much lower $Q$ and $F$, $\eta_1^{(fab)}$ decreases rapidly as the number of mirror periods becomes too low (Fig. 5c). Therefore, there is an optimal number of mirror periods that avoids too low reflectivity causing $Q$ and hence $F$ to lose its effectiveness in maintaining efficient $\eta_1$ but still allows efficient total outcoupling $\eta_{tot}$. We achieve an optimal design with $F = 10$ (with fab-imperfect $Q$) and $\eta_{tot} = 0.826$ at about 20 mirror periods. Once light propagates beyond the mirror region of the cavity, the diamond can be efficiently coupled to underlying SiN waveguides on an active photonics platform. Previous results have shown that optimized coupling between a tapered diamond waveguide and an underlying tapered SiN waveguide can reach up to 0.95 [43]. Combine this with SnV's high internal quantum efficiency of 0.8 [36], the total photon collection efficiency from the cavity to the waveguide is $\eta \approx 0.628$. This is well above collection efficiencies previously considered for calculating the percolation threshold of cluster states for quantum computing [44].

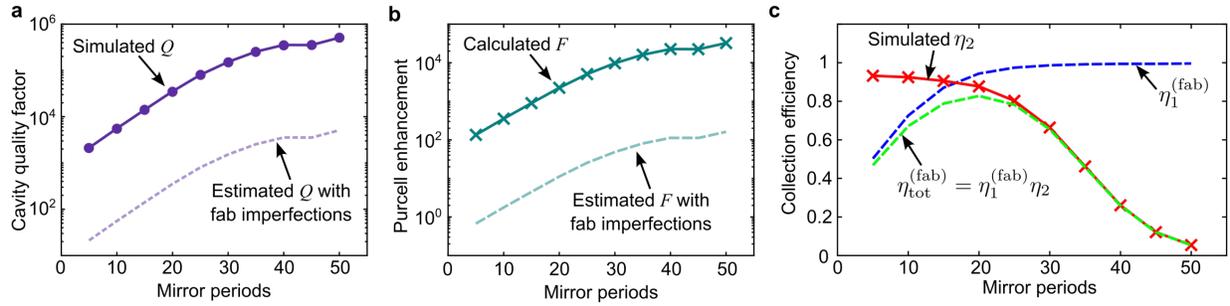

**Figure 5: Optimizing $\eta_{tot}$ for outcoupling of the hybrid cavity to a photonic integrated circuit.** Simulations performed for a dielectric pitch-defect cavity mode at 619.5 nm, with geometric parameters $a1 = 170$ nm, $a2 = 10$ nm, $a3 = 20$ periods, $a4 = 150$ nm, $a5 = 50\%$. $b1 = 340$ nm, $b4 = 100$ nm. a) Simulated $Q$ as the number of mirror periods on the photon collection side is varied. More mirror periods generally lead to higher $Q$. An estimated, lower $Q$ due to fab imperfections is also plotted. b) Calculated Purcell factor $F$ based on the simulated $Q$, including an estimated, lower $F$ due to fab imperfections. c) Coupling efficiencies $\eta_1^{(fab)}$, $\eta_2$, and $\eta_{tot}^{(fab)}$, which take into account the lower $Q$ and $F$ due to fab imperfections. There is an optimal mirror period number which balances high collection efficiency via the waveguide while maintaining a high Purcell factor of the cavity.

There are additional techniques that could be implemented to make outcoupling alignment simpler as well, such as with angled interconnects [45]. We note that vertical free-space collection designs are also possible, see Supplementary Section 1.

### Hybrid cavity and emitter tuning with piezoelectric actuation

While we presented an optical cavity with flexible characteristics, a further advantage of hybrid design is the ability to include CMOS-fabricated fine tuning of the underlying mirror dielectrics prior to hybrid cavity generation. This allows robust alignment of different color centers in different cavities all to a single frequency.

We propose a tuning method by attaching our hybrid cavity to a piezoelectric-piston body consisting of the same layer stack as our previously reported cryogenically compatible photonics technology [34], shown in Fig. 5. To calculate cavity mode tuning, we first modify the 619.5 nm cavity mode design to have only 20 mirror periods on each side (Fig. 5a). The mirror sections are etched through and connected only at the ends in a snake-like pattern, thereby reducing the overall mechanical rigidity of the cavity region. One side of the cavity is clamped while the other side is attached to a free-floating cantilever consisting of the piezoelectric layer (aluminum nitride) and electrodes (aluminum). The cantilever piston will readily push or pull against the cavity depending on the polarity of the applied voltage. By design, the generated strain will highly concentrate [46] in the cavity region due to the large difference between the effective mechanical compliance of the cantilever piston and the cavity mirror regions. To add a second tuning degree of freedom, an additional "ZPL" piezo is placed underneath the center of the cavity. By applying +/-100 V to this piezo, additional strain is generated in the diamond locally near the emitter of interest. The ZPL tuner allows us to control the color center's ZPL wavelength independently of the cavity piston tuner to better align with the optical resonance.

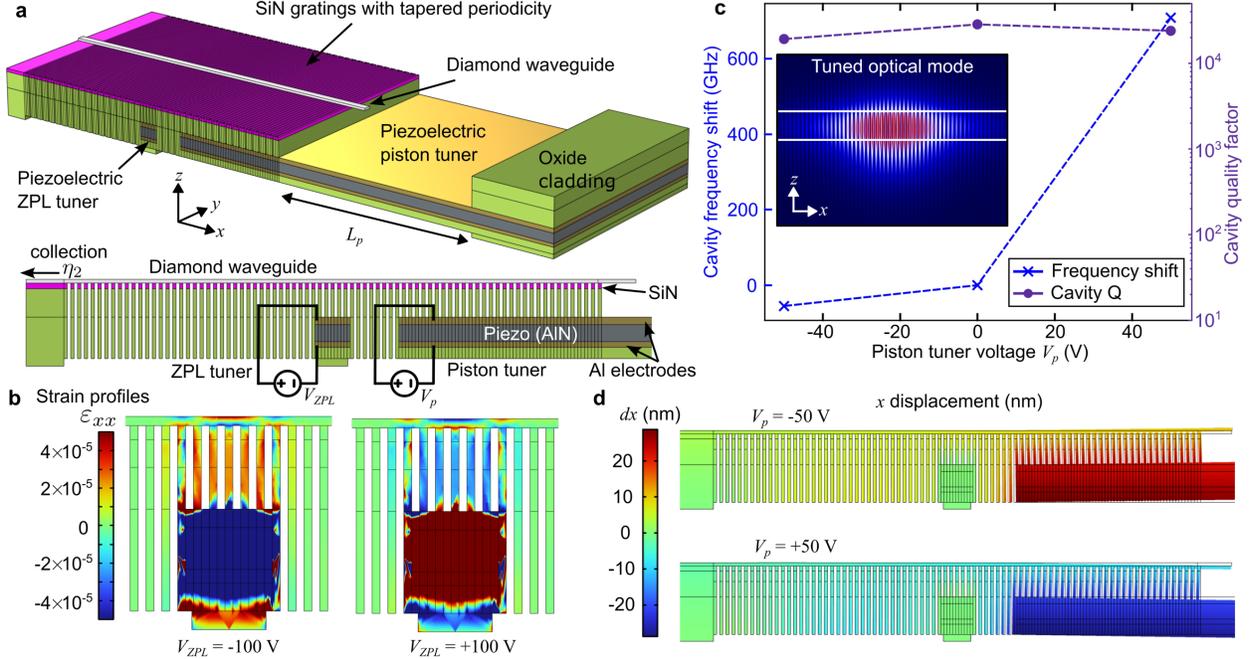

**Figure 6: Integrated piezoelectric tuning of the hybrid cavity mode.** a) Cavity strain tuning platform schematic showing the geometry of the hybrid cavity coupled to two independent piezoelectric tuners: ZPL tuner and piston tuner. Voltages $V_{ZPL}$ controls the ZPL tuner and $V_p$ the piston tuner. b) Strain profile of $\epsilon_{xx}$ when the ZPL tuner voltage is +/-100 V, showing strain propagating up to the diamond waveguide to tune the emitter's ZPL wavelength. c) Cavity mode (inset: E-field magnitude at $V_p = 0$, XZ cross section) tuning when applying voltage to the piston tuner, tuning up to 760 GHz from -50 to 50 V while preserving quality factor. d) Total x-displacement of the hybrid cavity when $V_p = +/-50$ V is applied to the piston cantilever.

We investigate the hybrid cavity tuning using finite-element (COMSOL) simulations, depicted in Fig. 6a. We first perform COMSOL simulations for a combination of piston (-50, 0, 50 V) and ZPL piezo (-100, 0, 100 V) voltages to strain the cavity. The re-meshed, deformed geometry and strain values are exported from COMSOL directly into Ansys Lumerical to accurately capture moving boundary contributions to the cavity mode shifts. For the strain-optic effects, we converted volume-averaged strains (Fig. 6b) to material index shifts in the deformed geometry, discretized into several regions. However, we found the strain-optic effect had an overall negligible effect on the cavity tuning (only <10 GHz shift, more details in Supplementary Section 2). The resulting cavity mode shifts based on primarily moving material boundaries are shown in Fig. 6c. We find that by applying voltages in the range of -50 to 50 V (readily accessible voltages on our platform), we can tune the cavity mode by ~750 GHz relative to the unperturbed mode. Fig. 6d shows the deformations induced by the piezoelectric piston at these voltages.

In this design, the ZPL of a color center in diamond will also be tuned by the strain field whose effect will vary somewhat depending on exact location in the cavity. Based on experimental strain tuning of SnVs with a ZPL near 619.5 nm of ~0.5 PHz/strain [41], our max strain values using the center ZPL piezo alone are on the order $10^{-5}$ (Fig. 6b) corresponding to a ZPL tuning of ~5 GHz. Given an implanted ensemble of SnVs with a transform-limited linewidth of ~30 MHz, we can readily tune across dozens of SnVs to isolate a single emitter within the cavity. We note there is additional tuning of the ZPL due to the strain induced by the piston cantilever, on the order $10^{-4}$ corresponding to a ~50 GHz shift for the ZPL. However, because the cavity mode shift is an order of magnitude larger, it is possible to find a pair of tuning voltages for the piston and ZPL tuner (fine adjustment) that align the cavity mode and the ZPL wavelengths. Lastly, since the strain due to the piston cantilever is approximately one order of magnitude higher than that due to the ZPL piezo, we found the cavity tuning resulting strictly from the ZPL piezo was below the accuracy threshold of the optical simulation.

**Discussion**

The hybrid cavity provides several inherent advantages in terms of fabrication reliability and scalability compared to existing approaches. Our approach transfers the complexity of high-resolution fabrication into mature CMOS materials such as SiN or $SiO_2$, enabling integration with existing complex photonic layer stacks [34] while increasing yield. These features would enable both direct piezoelectric tuning of the cavity (faster and more flexible than existing capacitive [47,48], laser tuning, [49,50] or gas tuning methods [9,51]) as well as photon collection into reconfigurable photonic circuits such as integrated switches and multiport interferometers [52]. On the diamond materials side, the geometry of the stamped waveguide may be post-selected for additional reliability when targeting a specific wavelength, such as 620 nm for SnV in diamond. Fabricating rectangular diamond waveguides is also a simpler process compared to fabricating fine-featured nanobeams or PhC cavities in the diamond itself. Additionally, the color centers may be placed further from diamond etched surfaces in the waveguide compared to a PhC, thereby decreasing potential effects of nearby surface states to cause defect ionization or spectral diffusion [53]. High quality fabrication of diamond waveguide micro-chiplets has already been demonstrated in prior works [40] and shown to couple efficiently to underlying PICs [43].

To realize our device experimentally, we may leverage our CMOS-fabricated photonics platform [34] by designing both the ZPL piezo and the piston tuner around a prepared film of SiN. We can utilize back-end-of-line processes such as electron beam lithography, focused ion beam, or deep UV lithography [54,55] to write the desired pitch-defect lines into the SiN. Once the SiN component is patterned and released, we can directly pick and place with a PDMS stamp to transfer a separately fabricated quantum diamond microchiplet (QMC) onto the SiN pattern to complete the hybrid cavity. The highly reflective end of the cavity is attached to the piston cantilever, whereas the end designed for photon output can have an array of underlying tapered SiN waveguides [41] where the output photons through the diamond waveguide can be coupled for further photon routing, entanglement, and detection schemes. The pick and stamp process may be scaled by potentially designing our wafer stack with a lock and release method previously demonstrated for large-scale heterogeneous integration of diamond on a CMOS platform [56].

**Funding**. MITRE Quantum Moonshot Project; DARPA ONISQ program; Brookhaven National Laboratory supported by U.S. Department of Energy, Office of Basic Energy Sciences, under Contract No. DE-SC0012704; NSF RAISE TAQS program; Center for Integrated Nanotechnologies, an Office of Science User Facility operated by the U.S. Department of Energy Office of Science.
**Acknowledgments**. M.D. and A.G. thank Y. H. Wen, G. Clark, D. A. Golter, K. J. Palm, and A. Khaykin for helpful technical discussions. M.D. and A.G. thank T. E. Espedal for providing auxiliary hybrid cavity simulations.
**Disclosures**. D.E. is a scientific advisor to and holds shares in QuEra Computing, Qunett, and Axiomatic_AI. The other authors declare no conflicts of interest.
**Data availability.** Data underlying the results presented in this paper are not publicly available at this time but may be obtained from the corresponding authors upon reasonable request.
**Supplemental document.** See Supplement for supporting content.

# Scalable construction of hybrid quantum photonic cavities: Supplementary Information

Andrew S. Greenspon,[1,2,7] Mark Dong,[1,2,8] Ian Christen,[2] Gerald Gilbert,[4] Matt Eichenfield,[3,5] and Dirk Englund[2,6,9]

[1]The MITRE Corporation, 202 Burlington Road, Bedford, Massachusetts 01730, USA
[2]Research Laboratory of Electronics, Massachusetts Institute of Technology, Cambridge, Massachusetts 02139, USA
[3]Sandia National Laboratories, P.O. Box 5800 Albuquerque, New Mexico, 87185, USA
[4]The MITRE Corporation, 200 Forrestal Road, Princeton, New Jersey 08540, USA
[5]College of Optical Sciences, University of Arizona, Tucson, Arizona 85719, USA
[6]Brookhaven National Laboratory, 98 Rochester St, Upton, New York 11973, USA
[7]agreenspon@mitre.org
[8]mdong@mitre.org
[9]englund@mit.edu

**Supplementary Section 1: Free space collection**

Collection efficiency $\eta_2$ can be measured for output via a waveguide or collection into a lens vertically above the cavity. Collection efficiency into the diamond waveguide on top of the hybrid cavity can be measured as the percent total emission confined in the waveguide mode as transmitted outside of the cavity region for routing light onto a PIC. For a cavity with infinite mirror sections, this value will be negligible. In order to improve transmission, the number of SiN nanostructured lines in the mirror section of the cavity can be decreased on one side until a large portion of the emission is output to this region. Decreasing the mirrors on one side will proportionally decrease $Q$ and a balance between $\eta_1$ and $\eta_2$ must be considered (Figure 5c).

Collection efficiency into a lens of a given numerical aperture (NA) above the cavity, or equivalently within a far-field divergence angle of $\theta_c = \sin^{-1}(NA)$ is

$$\eta_2 = T \left[ \int_0^{2\pi} d\phi \int_0^{\theta_c} d\theta \, |\vec{E}|^2 \sin(\theta) \right] / (\int_0^{2\pi} d\phi \int_0^{\pi} d\theta \, |\vec{E}|^2 \sin(\theta) ) \quad (S1)$$

where $|\vec{E}|^2$ is the far-field radiation intensity, $T$ is the percent of light emitted into the plane above the cavity relative to the total emission from the cavity in all directions. Optimizing the amount of light emitted within a smaller divergence angle may also decrease the $Q$ of the cavity from the Q-optimized design, though there are optimization methods to balance Purcell enhancement and $\eta_2$ to get the highest figure of merit (FoM) $\eta_{tot}$ [39].

To design a cavity for vertical outcoupling, we add a grating perturbation on top of the pitch taper of the dielectric cavity. Using a basic grating perturbation within the cavity region leads to a focus of the output of the cavity toward a narrower far-field angle upward for collection with an objective for potential photon routing in free space. Further optimization protocols such as via gradient descent or more complex algorithms should be able to achieve even higher vertical outcoupling efficiency with higher quality factors [39,49]. Last, we add a metal backplane below the oxide to decrease the amount of light that is lost to the substrate. Using this first order method, we couple up to 58.7% of all emitter light into an objective with NA = 0.5 and 72.5% into an objective with NA = 0.9. Theoretical quality factor in this case is 4100 and can be further optimized.

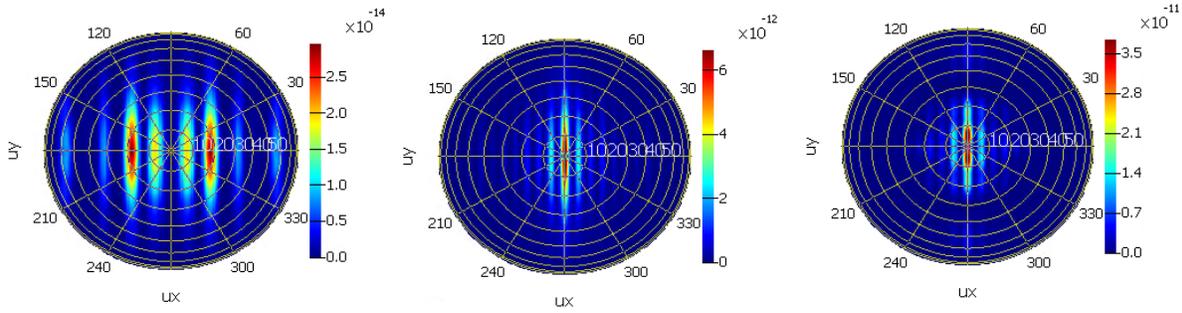

**Figure S1. Improving output of the cavity free space collection vertically upward for dielectric pitch defect cavity.** Geometric parameters: $a1 = 170$ nm, $a2 = 10$ nm, $a3 = 20$ periods, $a4 = 150$ nm, $a5 = 50\%$. $b1 = 340$ nm, $b2 = b3 = 0$, $b4 = 100$ nm. Cavity mode ~619.5 nm. Total transmission into an objective with NA = 0.5 or 0.9. Far-field profile of the magnitude of the electric field squared for: left) $Q$-optimized structure, $T$ (NA = 0.5) = 9.7%, $T$ (NA = 0.9) = 16.7%. center) With a grating perturbation super-imposed on the quadratic periodic taper of the underlying SiN nanolines, showing increased E-field propagation vertically upward and focused within a narrower cone for collection. $T$ (NA = 0.5) = 15.7%, $T$ (NA = 0.9) = 23.0%. right) for the same cavity as in center)., but with a metal backplane at 650 nm below the bottom of the SiN lines. $T$ (NA = 0.5) = 54.6%, $T$ (NA = 0.9) = 71.2.0%.

## Supplementary Section 2: Cavity mode shift due to variation in etch duty cycle

Variation in the etch duty cycle (fill factor of the SiN gratings) shifts the cavity mode, but the quality factor remains high. We simulated the pitch-defect design with $a1 = 170$ nm, $a2 = 10$ nm, $a3 = 20$ periods, $a4 = 150$ nm, $b4 = 100$ nm, and varied $a5$ at 40%, 50%, 60% duty cycle. Each of these simulations were run for 2000 fs and quality factor was calculated via the exponential decay of the electric field envelope. Etch duty cycles of 40% and 60% for the 340 nm wide diamond, shift the cavity mode to 625 and 613 nm respectively, but quality factor remains $Q > 10^5$. Mode shape is qualitatively the same, for single TE mode coupling to an underlying PIC. To be within the tuning range of the piston cantilever, variation between SiN gratings on a given reticle should be less than ~3.3%. Pitch tolerance for 3.3% for 170 nm center pitch would give 5.6 nm pitch tolerance and 2.8 nm SiN line tolerance for a 50% duty cycle pattern.

If the as fabricated etch duty cycle varied up to 10% relative to the design, the cavity mode would shift ~ 5 nm for the same diamond waveguide geometry. From Figure 4, we see that we can coarsely tune the cavity back to the desired cavity mode by adjusting the diamond waveguide width by ~20 nm. These tolerances should be achievable for diamond waveguide fabrication. Fine tuning of the cavity mode can then be performed using the piston and ZPL cantilevers.

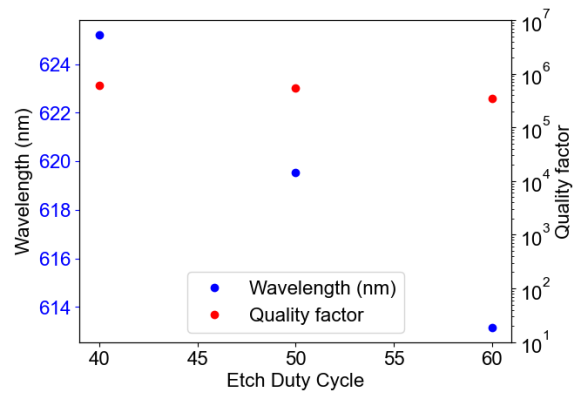

**Figure S2. Cavity mode tolerance to variations in etch duty cycle**

**Supplementary Section 3: Strain-optic effect in cavity tuning 3D simulations**

We performed the piezoelectric cavity tuning simulations using a combination of commercial finite-element software. Specifically, we simulated the piezo-mechanical coupling on COMSOL Multiphysics and the optical cavity modes on ANSYS Lumerical. In order to account for all relevant mechanical effects from the piezoelectric actuation, we exported the remeshed deformed geometry from COMSOL (in the form of a binary .stl file) and imported directly into Lumerical for the optical simulations. This approach takes advantage of the powerful optical time-domain techniques in Lumerical while leveraging efficient piezoelectric-MEMS multiphysics simulations in COMSOL.

To further improve the simulation accuracy, we also incorporated refractive index changes due to strain-optic effects in the diamond and silicon nitride layers. We estimated the refractive index shifts as follows.

The change in index ellipsoid coefficients $\Delta B$ for a cubic or isotropic material can be expressed as [57]

$$\begin{bmatrix} \Delta B_1 \\ \Delta B_2 \\ \Delta B_3 \\ \Delta B_4 \\ \Delta B_5 \\ \Delta B_6 \end{bmatrix} = \begin{bmatrix} p_{11} & p_{12} & p_{12} & 0 & 0 & 0 \\ p_{12} & p_{11} & p_{12} & 0 & 0 & 0 \\ p_{12} & p_{12} & p_{11} & 0 & 0 & 0 \\ 0 & 0 & 0 & p_{44} & 0 & 0 \\ 0 & 0 & 0 & 0 & p_{44} & 0 \\ 0 & 0 & 0 & 0 & 0 & p_{44} \end{bmatrix} \begin{bmatrix} \epsilon_1 \\ \epsilon_2 \\ \epsilon_3 \\ \epsilon_4 \\ \epsilon_5 \\ \epsilon_6 \end{bmatrix} \quad (S2)$$

where $p_{ij}$ are the strain-optic coefficients and $\epsilon_j$ is the linear strain tensor, both in Voigt notation. We defined the propagation direction as the x axis and nominally the TE polarization direction as the y axis. In this case, the relevant index shifts reduce to just the $\Delta B_2 = \Delta B_{yy}$ term. To first order, we have ignored the non-principal strains. Taylor expansion of $\Delta B_2$ around $\Delta n_y = 0$ yields the simplified equation

$$\Delta B_{yy} = \frac{1}{(n_0 + \Delta n_y)^2} - \frac{1}{n_0^2} \approx -\frac{2}{n_0^3} \Delta n_y \quad (S3)$$

$$\Delta n_y = -\frac{n_0^3}{2}(p_{11}\epsilon_{yy} + p_{12}(\epsilon_{xx} + \epsilon_{zz})) \quad (S4)$$

We used the values of $p_{11} = -0.249$, $p_{12} = 0.043$ [58] for diamond and $p_{11} = 0.08$, $p_{12} = 0.047$ [59] for silicon nitride. We note the silicon nitride strain-optic coefficients are best estimates and not well established in the literature [60]. However, in the end, we found the strain-optic contribution to the cavity mode shift (between 1 GHz - 10 GHz, close to the simulation accuracy) was negligible compared to the geometric shift which accounts for almost all of the 760 GHz tuning range.